\documentclass{webofc}
\usepackage[varg]{txfonts}   
\usepackage{bm}
\usepackage{bbm}
\begin{document}
\title{
Strange hidden-charm pentaquark poles from
$B^-\to J/\psi\Lambda\bar{p}$ 
}
%
%

\author{\firstname{Satoshi X.} \lastname{Nakamura}\inst{1}\fnsep\thanks{\email{satoshi@sdu.edu.cn}} \and
        \firstname{Jia-Jun} \lastname{Wu}\inst{2}
 \fnsep\thanks{\email{wujiajun@ucas.ac.cn}}
}

\institute{Institute of Frontier and Interdisciplinary Science, Shandong
University, Qingdao 266237, China
\and
School of Physical Sciences, University of Chinese Academy of Sciences, Beijing 100049, China
          }

\abstract{%
Recent LHCb data for $B^-\to J/\psi\Lambda\bar{p}$ 
show a clear peak structure at the $\Xi_c\bar{D}$
threshold in the $J/\psi\Lambda$ invariant mass ($M_{J/\psi\Lambda}$)
distribution.
The LHCb's amplitude analysis identified the peak with 
the first hidden-charm pentaquark with strangeness
$P_{\psi s}^\Lambda(4338)$.
We conduct a coupled-channel amplitude analysis of the LHCb data by 
simultaneously fitting the
$M_{J/\psi\Lambda}$, $M_{J/\psi\bar{p}}$, $M_{\Lambda\bar{p}}$, and
$\cos\theta_{K^*}$ distributions.
Rather than the Breit-Wigner fit employed in the
 LHCb analysis,
we consider relevant threshold effects and 
 a unitary $\Xi_c\bar{D}$-$\Lambda_c\bar{D}_s$ coupled-channel
 scattering amplitude from which $P_{\psi s}^\Lambda$ poles are extracted 
 for the first time. 
In our default fit, 
the $P_{\psi s}^\Lambda(4338)$ pole is almost a
$\Xi_c \bar{D}$ bound state at
$( 4338.2\pm 1.4)-( 1.9\pm 0.5 )\,i$~MeV.
Our default model also fits
a large fluctuation at
the $\Lambda_c\bar{D}_s$ threshold, giving
a $\Lambda_c\bar{D}_s$ virtual state, $P_{\psi s}^\Lambda(4255)$, 
at $4254.7\pm 0.4$~MeV.
We also found that 
the $P_{\psi s}^\Lambda(4338)$ peak cannot solely be a kinematical
effect, and a nearby pole is needed.
}
\maketitle
\section{Introduction}

The first discovery of a strange hidden-charm pentaquark $P_{\psi s}^\Lambda(4338)$
in $B^-\to J/\psi\Lambda\bar{p}$
was recently reported by the LHCb Collaboration~\cite{lhcb_seminar}.
The data shows a clear peak in the $J/\psi\Lambda$ invariant mass 
($M_{J/\psi\Lambda}$) distribution, suggesting the pentaquark contribution.
The LHCb conducted an amplitude analysis to determine
the pentaquark mass, width, and spin-parity to be
$ 4338.2\pm 0.7~{\rm MeV}$, $ 7.0\pm 1.2~{\rm MeV}$, and $J^P=1/2^-$, respectively.
These resonance parameters are primary basis to address 
the nature and internal structure of $P_{\psi s}^\Lambda(4338)$.
Our concern here is that the LHCb analysis assumed that
the resonance-like peak can be well described with 
a Breit-Wigner (BW) amplitude.
Actually, the resonance peak sits right on the 
$\Xi_c\bar{D}$ threshold [see Fig.~\ref{fig:comp-data}(a)].
The BW fit is often unsuitable in this situation 
because a kinematical effect (threshold cusp) 
may cause the resonancelike structure.
Even if there exists a relevant pole that couples with the
 $\Xi_c\bar{D}$ channel, 
 the branch cut from the $\Xi_c\bar{D}$ channel would 
distort the lineshape due to the pole,
invalidating the BW fit. 

Thus, in this work, we conduct a coupled-channel amplitude analysis 
of the LHCb data on  $B^-\to J/\psi\Lambda\bar{p}$
with all relevant kinematical effects taken into account;
see Ref.~\cite{ours} for details.
We fit our amplitude model to 
the $M_{J/\psi\Lambda}$, $M_{J/\psi\bar{p}}$, $M_{\Lambda\bar{p}}$, and
$\cos\theta_{K^*}$
distribution data simultaneously.
Our model does not include BW amplitudes but
a unitary $\Xi_c\bar{D}$-$\Lambda_c\bar{D}_s$ coupled-channel amplitude
with which we address whether the LHCb data requires pentaquark poles. 

\begin{figure*}[t]
\begin{center}
\includegraphics[width=1\textwidth]{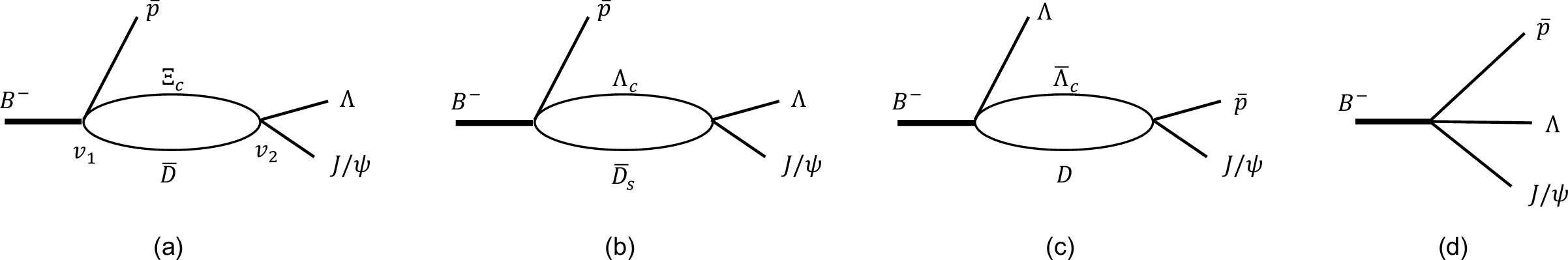}
\end{center}
 \caption{
$B^-\to J/\psi \Lambda \bar{p}$ mechanisms 
considered in this work.
The diagrams (a)-(d) have different 
weak vertices $v_1$ such as
(a) $B^-\to\Xi_c\bar{D}\bar{p}$,
(b) $B^-\to\Lambda_c\bar{D}_s\bar{p}$,
(c) $B^-\to\bar{\Lambda}_c D\Lambda$, and 
(d) $B^-\to J/\psi\Lambda\bar{p}$.
The vertex $v_2$ in (a,b)
is a $\Xi_c\bar{D}-\Lambda_c\bar{D}_s$ coupled-channel 
scattering and a perturbative transition to
$J/\psi\Lambda$ while, in (c), 
an elastic $\bar{\Lambda}_c D$ scattering
and  a perturbative transition to $J/\psi\bar{p}$.
Figures taken from Ref.~\cite{ours}. Copyright (2023) APS.
 }
\label{fig:diag}
\end{figure*}

\section{Model}

In the invariant mass distributions of
$B^-\to J/\psi\Lambda\bar{p}$,
noticeable structures can be seen at 
the $\Xi_c\bar{D}$, $\Lambda_c\bar{D}_s$, and 
$\bar{\Lambda}_c D$ thresholds.
This suggests that 
threshold cusps from the diagrams in Figs.~\ref{fig:diag}(a-c) cause the structures;
hadronic rescatterings and the
associated poles could 
further enhance or suppress the cusps.
Thus our amplitude model considers
the diagrams of Figs.~\ref{fig:diag}(a-c),
and also a direct decay of
Fig.~\ref{fig:diag}(d) that would absorb
other possible mechanisms.
We consider only $s$-wave interactions that are expected to be
dominant since the $Q$-value is not so large ($\sim 130$~MeV).

We include the most important coupled-channels in the hadronic scatterings;
a $\Xi_c\bar{D}-\Lambda_c\bar{D}_s(1/2^-)$ coupled-channel in
Figs.~\ref{fig:diag}(a,b), and a $\bar{\Lambda}_c D(1/2^+)$
single-channel in Fig.~\ref{fig:diag}(c).
Our data-driven approach employs contact separable 
hadron interactions not biased by any particular
models, and 
determine all coupling strengths by fitting the data.
The relevant coupled-channel unitarity is respected.
These scatterings are followed by perturbative 
transitions to the final $J/\psi\Lambda$ and $J/\psi\bar{p}$ states in
our model.

\section{Results}
\begin{figure*}[t]
\begin{center}
\includegraphics[width=1\textwidth]{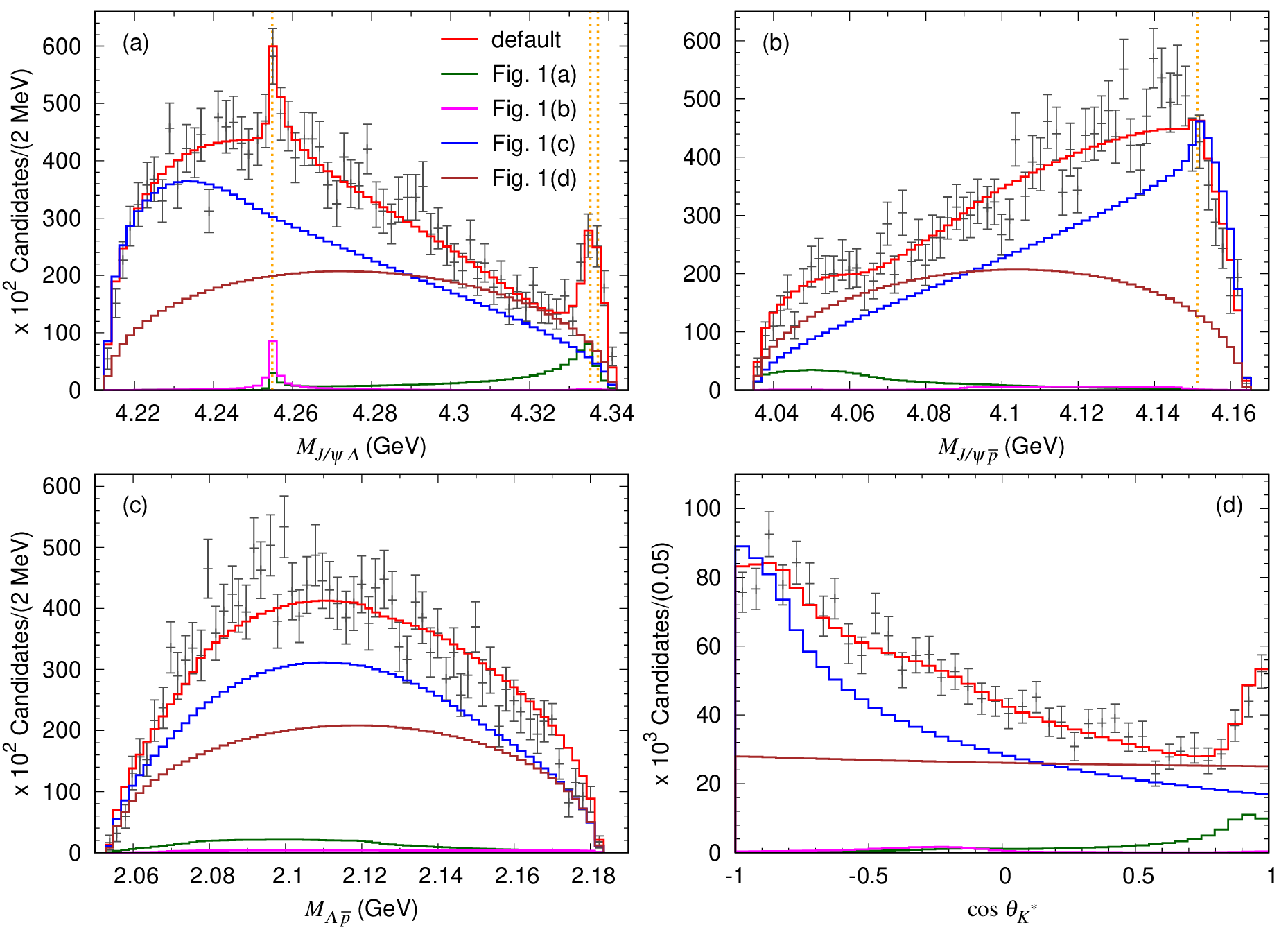}
\end{center}
 \caption{
Simultaneous fit to
(a) $J/\psi \Lambda$, (b) $J/\psi \bar{p}$,
(c) $\Lambda \bar{p}$ invariant mass,
and (d) $\cos\theta_{K^*}$ distributions
  of $B^-\to J/\psi \Lambda \bar{p}$ from
 the LHCb~\cite{lhcb_seminar} ;
efficiency-corrected and background-subtracted data.
The plots are 
the default fit and 
each contribution from diagrams in Fig.~\ref{fig:diag}.
The dotted vertical lines 
in the panel (a) [(b)] indicate
$\Lambda_c^+ D_s^-$, 
$\Xi_c^0 \bar{D}^{0}$, and
$\Xi_c^+ D^-$ 
[$\bar{\Lambda}_c^- D^0$]
thresholds
from left to right.
Figures taken from Ref.~\cite{ours}. Copyright (2023) APS.
 }
\label{fig:comp-data}
\end{figure*}
The $M_{J/\psi\Lambda}$, $M_{J/\psi\bar{p}}$, $M_{\Lambda\bar{p}}$, and
$\cos\theta_{K^*}$ distributions
from the LHCb are simultaneously fitted with our model described in
the previous section;
$\cos\theta_{K^*}\equiv \bm{p}_\Lambda\cdot \bm{p}_\psi / |\bm{p}_\Lambda||\bm{p}_\psi|$ 
in the $\Lambda\bar{p}$ center-of-mass frame.
In our default fit, we adjust 9 fitting parameters from 
coupling strengths of the weak vertices and hadronic interactions.
The fit result is shown in Fig.~\ref{fig:comp-data};
$\chi^2/{\rm ndf}\simeq 1.21$ with
'ndf' being the number of bins minus the number of the fitting parameters.
The presented binned theoretical distributions are obtained by
smearing
theoretical invariant mass ($\cos\theta_{K^*}$)
distributions with experimental resolutions of 1~MeV 
(bin width of 0.05), and then averaging them over the bin width in each
bin. 
The LHCb data, including the $P_{\psi s}^\Lambda(4338)$ peak at
$M_{J/\psi\Lambda}\sim 4338$~MeV,
are well fitted by our default model as seen in 
Fig.~\ref{fig:comp-data}.
Our default model also fits 
a large fluctuation at $M_{J/\psi\Lambda}\sim 4255$~MeV.
The LHCb analysis concluded this fluctuation to be
a statistical one.
However, considering the fact that 
the fluctuation sits just right on
the $\Lambda_c \bar{D}_s$ threshold, 
we can expect a visible threshold cusp
from a color-favored $B^-\to \Lambda_c \bar{D}_s \bar{p}$ followed by 
 $\Lambda_c \bar{D}_s\to J/\psi \Lambda$.
The cusp might have been enhanced by 
a $\Lambda_c \bar{D}_s$ rescattering and an associated 
$P_{\psi s}^\Lambda(4255)$ pole.

Each Contribution from the diagrams in Fig.~\ref{fig:diag} 
is also given in Fig.~\ref{fig:comp-data}.
Dominant mechanisms are 
Figs.~\ref{fig:diag}(c) [blue] and \ref{fig:diag}(d) [brown].
We can understand that
the increasing $M_{J/\psi\bar{p}}$ distribution
in Fig.~\ref{fig:comp-data}(b) 
is from Fig.~\ref{fig:diag}(c) that causes the $\bar{\Lambda}_c D$
threshold cusp.
Our fit found that 
the cusp is suppressed by
a repulsive $\bar{\Lambda}_c D$ interaction, which is consistent
with our previous analysis of $B^0_s\to J/\psi p\bar{p}$~\cite{sxn_Bs}.
Contributions from
the diagrams of Figs.~\ref{fig:diag}(a) [green] and
\ref{fig:diag}(b) [magenta] are smaller in the magnitude.
However, they show
significantly enhanced $\Xi_c\bar{D}$ and $\Lambda_c\bar{D}_s$
threshold cusps.
The $P_{\psi s}^\Lambda$ peaks are caused by them through the interference. 

There are qualitative differences between our and LHCb's descriptions of
the data.
In the LHCb analysis, the $M_{J/\psi\bar{p}}$ distribution is fitted with
a non-resonant 
$p$-wave $J/\psi\bar{p}$ [NR($J/\psi\bar{p}$)] amplitude in a
polynomial form, and the physical origin of the increasing
behavior is not clarified. 
The NR($J/\psi\bar{p}$) contribution reaches $\sim 84$\% fit fraction.
Since a $s$-wave dominance is usually expected in 
the small $Q$-value process,
this $p$-wave dominance is difficult to understand.
Our model includes $s$-wave $J/\psi\bar{p}$ only.
Regarding the number of fitting parameters, 
16 in the LHCb's model while 
9(8) in our default (alternative) model.
Since the LHCb fitted richer information from six-dimensional data, 
they would need more parameters.
However, 
this might not fully explain
$\sim 2$ times more parameters.
Rather, we suspect that 
the $p$-wave dominance and excessive parameters are due to
missing relevant mechanisms such as Figs.~\ref{fig:diag}(a-c),
since many other mechanisms would be needed to mimic the relevant ones
through complicated interferences.

\begin{figure}[t]
\begin{center}
\includegraphics[width=.5\textwidth]{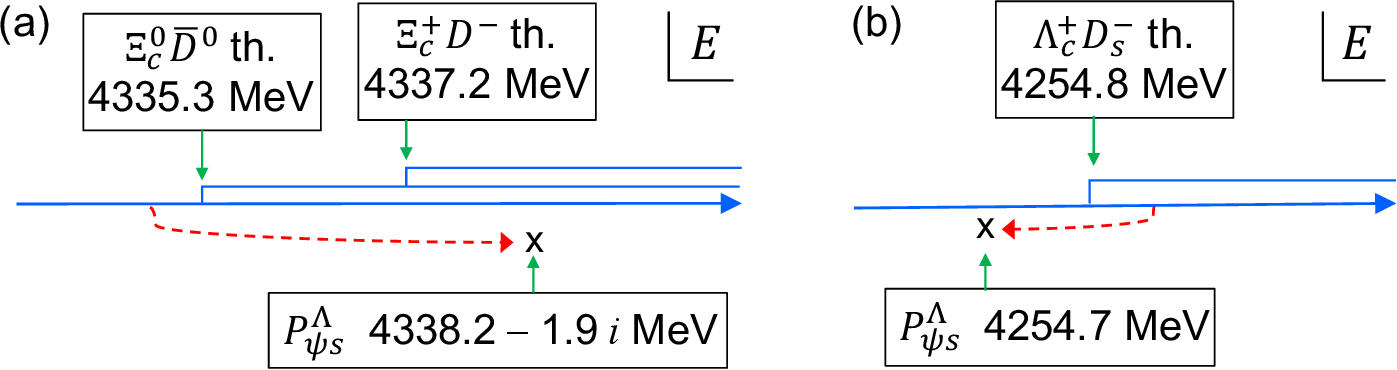}
\end{center}
 \caption{
(a)[(b)] $P_{\psi s}^\Lambda(4338)$ [$P_{\psi s}^\Lambda(4255)$] pole
 position from the default model. 
The branch cuts are indicated by the double lines.
The red dotted arrows connect the poles and their closest physical energy regions.
Figures taken from Ref.~\cite{ours}. Copyright (2023) APS.
 }
 \label{fig:pole_sheet}
\end{figure}

Our default
$\Xi_c\bar{D}-\Lambda_c\bar{D}_s(1/2^-)$ coupled-channel scattering
amplitude is analytically continued to find relevant poles.
We found $P_{\psi s}^\Lambda(4338)$ pole at 
$( 4338.2\pm 1.4)-( 1.9\pm 0.5 )\,i$~MeV;
$J^P=1/2^-$ is consistent with the LHCb result.
We also found
$P_{\psi s}^\Lambda(4255)$ pole at
$ 4254.7\pm 0.4$~MeV.
Figure~\ref{fig:pole_sheet} illustrates where 
the poles are located relative to the relevant thresholds.
As seen in the figure,
the $P_{\psi s}^\Lambda(4338)$ pole is 
a $\Xi_c \bar{D}$ bound state slightly shifted due to a coupled-channel
effect.
Also, 
the $P_{\psi s}^\Lambda(4255)$ pole is essentially
a $\Lambda_c\bar{D}_s$ virtual state.

\begin{figure}[b]
\begin{center}
\includegraphics[width=.5\textwidth]{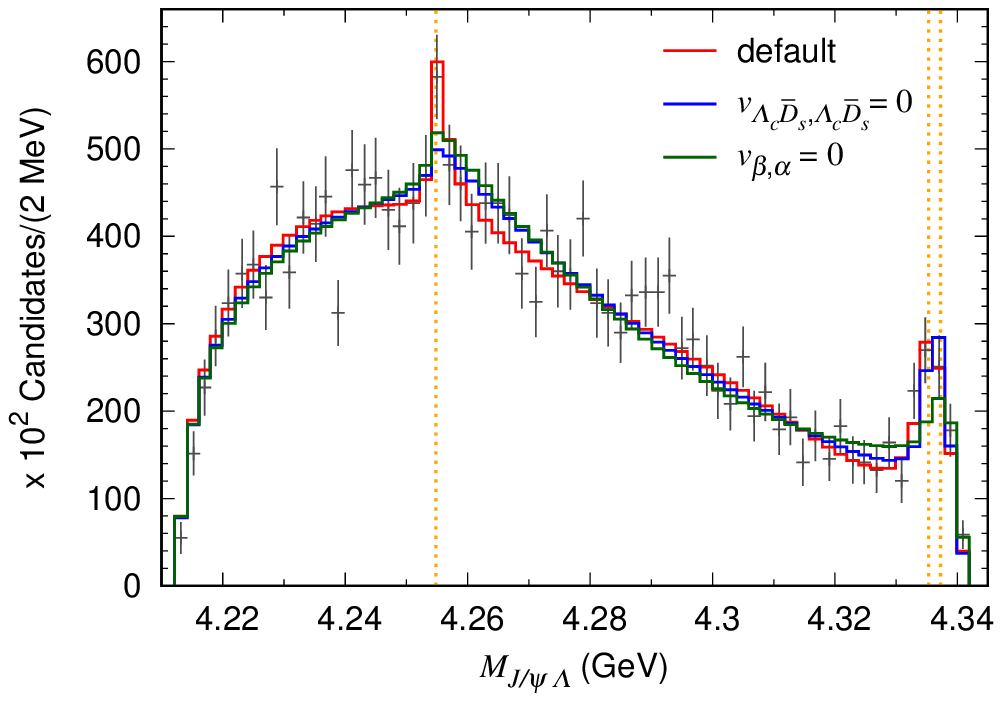}
\end{center}
 \caption{
Default and alternative fits to the LHCb data
  for $B^-\to J/\psi \Lambda \bar{p}$~\cite{lhcb_seminar}.
Figures taken from Ref.~\cite{ours}. Copyright (2023) APS.
 }
\label{fig:comp-models}
\end{figure}

We also considered alternative models without
$P_{\psi s}^\Lambda(4255)$ pole, and with/without energy dependence in
$\Xi_c\bar{D}$ interaction.
We obtained comparable fits 
as shown in Fig.~\ref{fig:comp-models}[blue].
There is still a $\Lambda_c\bar{D}_s$ threshold cusp without a nearby pole.
The default and alternative models have $P_{\psi s}^\Lambda(4338)$
poles on different Riemann sheets, suggesting the need of more precise data
for $B^-\to \Lambda_c \bar{D}_s \bar{p}$ and also
$\Xi^-_b \to J/\psi\Lambda K^-$.
We also examined if the $\Xi_c\bar{D}$ threshold cusp without a nearby pole
can explain the $P_{\psi s}^\Lambda(4338)$ peak, as shown in 
Fig.~\ref{fig:comp-models} [green].
We find a noticeably worse fit in the $P_{\psi s}^\Lambda(4338)$ region,
concluding that 
a nearby pole is needed to enhance the cusp.\\

{\bf\noindent Acknowledgments}\\
This work is in part supported by 
National Natural Science Foundation of China (NSFC) under contracts 
U2032103 (S.X.N.) and under Grants No. 12175239 and 12221005 (J.J.W.).


\begin{thebibliography}{}

\bibitem{lhcb_seminar}
R. Aaij et al. (LHCb Collaboration),
Phys. Rev. Lett. {\bf 131}, 031901 (2023).

 \bibitem{ours}
S.X. Nakamura and J.-J. Wu,
Phys. Rev. D {\bf 108}, L011501 (2023).

\bibitem{sxn_Bs}
S.X. Nakamura, A. Hosaka, and Y. Yamaguchi,
Phys. Rev. D {\bf 104}, L091503 (2021).


\end{thebibliography}


\end{document}